\documentclass[epj]{svjour}
\usepackage[utf8]{inputenc}
\usepackage[T1]{fontenc}
\usepackage{graphics}
\usepackage{graphicx}
\usepackage{amsmath}
\usepackage{gensymb}
\usepackage{soul}
\usepackage{color}

\begin{document}
\title{Equilibrium and non-equilibrium concentration fluctuations in a critical binary mixture}
\author{Fabio Giavazzi\inst{1} \and Alessandro Fornasieri\inst{1} \and Alberto Vailati\inst{2} \and Roberto Cerbino\inst{1}
\thanks{\emph{email:} roberto.cerbino@unimi.it}%
}                     
\offprints{}          
\institute{Dipartimento di Biotecnologie Mediche e Medicina Traslazionale, Università degli Studi di Milano, via F.lli Cervi 93, 20090 Segrate, Italy \and Dipartimento di Fisica, Università degli Studi di Milano, via Celoria 16, 20133 Milano, Italy}
\date{Received: date / Revised version: date}
%
\abstract{
When a macroscopic concentration gradient is present across a binary mixture, long-ranged \textit{non-equilibrium} concentration fluctuations (NCF) appear as a consequence of the coupling between the gradient and spontaneous equilibrium velocity fluctuations. Long-ranged \textit{equilibrium} concentration fluctuations (ECF) may be also observed when the mixture is close to a critical point. Here we study the interplay between NCF and critical ECF in a near critical mixture aniline/cyclohexane in the presence of a vertical concentration gradient. To this aim, we exploit a commercial optical microscope and a simple, custom-made, temperature-controlled cell to obtain simultaneous static and dynamic scattering information on the fluctuations. We first characterise the critical ECF at fixed temperature $T$ above the upper critical solution temperature $T_{c}$, in the wide temperature range $T-T_{c}\in[0.1,30]$ $^{o}$C. In this range, we observe the expected critical scaling behaviour for both the scattering intensity and the mass diffusion coefficient and we determine the critical exponents $\gamma$, $\nu$ and $\eta$, which are found in agreement with the 3D Ising values. We then study the system in the two-phase region ($T<T_{c}$). In particular, we characterise the interplay between ECF and NCF when the mixture, initially at a temperature $T_{i}$, is rapidly brought to a temperature $T_{f}>T_{i}$. During the transient, a vertical diffusive mass flux is present that causes the onset of NCF, whose amplitude vanishes with time, as the flux goes to zero. We also study the time dependence of the equilibrium scattering intensity $I_{eq}$, of the crossover wave-vector $q_{co}$ and of the diffusion coefficient $D$ during diffusion and find that all these quantities exhibit an exponential relaxation enslaved to the diffusive kinetics.
\PACS{
      {PACS-key}{describing text of that key}   \and
      {PACS-key}{describing text of that key}
     } 
} 
\maketitle

\section*{Introduction}

When a binary mixture prepared at the critical composition is brought close to the critical temperature $T_{c}$, a dramatic enhancement of the spatial correlation length of concentration fluctuations and an impressive slowing down of their diffusive relaxation dynamics can be observed, for instance with scattering experiments \cite{criticalbook,Peliti:2011rt}. Experiments with a binary mixture are very convenient for probing the universal properties of second-order phase transitions: mixtures are simple to handle and prepare at atmospheric pressure and often exhibit $T_{c}$ that are close enough to room temperature. For quite similar reasons, binary critical mixtures have been also shown to be ideal samples for the investigation of the so called non-equilibrium fluctuations i.e. the long-ranged fluctuations occurring in fluids that are kept out of equilibrium by an external gradient, for instance of temperature or concentration \cite{De-Zarate:2006qe}. In particular, when a critical binary mixture initially in the two phase region is rapidly brought in the one-phase region by a suitable temperature jump, the sharp concentration gradient present within the mixture gives rise to diffusion and amplifies spontaneous equilibrium velocity fluctuations by turning them into spectacularly large (a.k.a. giant) NCF \cite{Vailati:1997kq,Cicuta:2000cr,Cicuta:2001nx}. While in the absence of gravity these NCF relax diffusively as in equilibrium and their size can become as large as the sample container, on Earth the situation is different: both the amplitude and the lifetime of NCF are in fact reduced by gravity, which causes large fluctuations to relax by buoyancy rather than by diffusion \cite{Segre:1993qd,Vailati:2006yq,Vailati:2011qf,Cerbino:2015rt,Croccolo:2016mz}. A similar amplification/quenching mechanism of fluctuations occurs in the case of non-equilibrium temperature fluctuations induced by the presence of a temperature gradient \cite{takacs11}.

In this work, we study the interplay between NCF and ECF in the critical mixture aniline-cyclohexane at different temperatures above and below the upper critical solution temperature $T_{c}$. Our study addresses simultaneously the structure and the dynamics of the fluctuations, which are quantified by means of Differential Dynamic Microscopy (DDM) \cite{Cerbino:2008if,Giavazzi:2009xd}, a recently proposed Digital Fourier Microscopy method \cite{Giavazzi:2014it} that makes use of a commercial microscope equipped with a digital camera. We also use a novel custom-made cell specifically developed for guaranteeing easy optical access with the microscope, accurate and stable temperature control and a rapid sample thermalisation.\\ We first validate our approach by assessing the scaling properties observed when the mixture, prepared at the critical composition, is brought close to $T_{c}$ in the one phase region. In the wide temperature range $T-T_{c}\in[0.1,30]$ $^{o}$ C, our results for the scattering intensity and the diffusion coefficient as a function of temperature show an excellent agreement with the most accurate values available for the critical exponents for the 3D Ising universality class \cite{Pelissetto:2002cs}, which demonstrates the accuracy and sensitivity of our setup.

We then studied the process by which a mixture in equilibrium at a temperature $T_{i}<T_{c}$ finds its new equilibrium state when the sample temperature is suddenly changed to $T_{f}>T_{i}$, with $T_{f}<T_{c}$. We find that the transition between these two states (in both of which the sample is separated in two phases) is characterised by a monotonic decrease of the diffusion coefficient and a monotonic increase of the scattering intensity of ECF that are both well described by an exponential relaxation with time constant of about 50 min, compatible with the diffusive relaxation of the overall concentration profile across the cell. 
Immediately after the temperature switch large NCF develop, whose amplitude monotonically decreases in time. The value of the so-called cross-over wave-vector $q_{co}$, which identifies the smallest length-scale for which the amplitude of NCF is larger than its equilibrium counterpart, is found to vanish in time with the same relaxation kinetics of the concentration gradient at the interface between the two phases, in agreement with theory.

\section*{Theory: concentration fluctuations in a near-critical binary mixture}\label{theory}
In this work, we investigate the interplay between ECF, which are always present in a binary mixture due to thermal fluctuations, and NCF, which arise the presence of a macroscopic concentration gradient across the mixture. The spatial correlations of the fluctuations are described by their \textit{static structure factor}, which is defined as $S(\mathbf{q})=\left\langle \left|\delta c(\mathbf{q},t)\right|^{2}\right\rangle _{t}$. Here $c(\mathbf{x},t)$ is the weight fraction of the denser component of the mixture, $t$ is time, and $\mathbf{q}$ is the wave-vector of the fluctuations, the conjugated variable to the position vector $\mathbf{x}$ in a Fourier transform operation. In all the cases studied in this work the structure factor can be written as \begin{equation}
  S(\mathbf{q})=S_{eq}(\mathbf{q})+S_{ne}(\mathbf{q})
  .\label{sommaS}
\end{equation}
 The first term on the right hand side (rhs) Eq. \ref{sommaS} represents the equilibrium contribution to the static factor, whereas the second one describes the non-equilibrium excess due to the presence of a macroscopic concentration gradient inside the mixture. In the rest of this Section we will give explicit expressions for these two terms, as well as for the relevant transport coefficients, which can be extracted from scattering experiments \cite{Berne:1976yu}.

\subsection*{Equilibrium concentration fluctuations (ECF)}    
\label{ecf}

ECF in a binary liquid mixture arise as a consequence of the random thermal motion of molecules. The local changes in the concentration determine in turn fluctuations of the index of refraction that can be probed by using light scattering techniques \cite{mountain69}. The experimental investigation of equilibrium fluctuations in molecular mixtures is in general not trivial, due to the fact that they have usually a quite small amplitude and a very fast dynamics. Fluctuations with larger amplitude and slower dynamics can be obtained by using a near critical mixture i.e. a mixture that exhibits a second order phase transition between a one-phase region, in which two components coexist in a single phase, and a two-phase region, in which the mixture separates into two liquid phases \cite{criticalbook}. Such a system is characterised by the presence of a \textit{critical consolution point}, close to which point a small change in the temperature can lead to the separation of the mixture into two distinct liquid phases. When approaching the critical point, the correlation length and the amplitude of the fluctuations diverge, thus permitting an easier detection of the concentration fluctuations with respect to the case of molecular mixtures far from a critical point.\\In the following we will consider a near-critical binary liquid mixture prepared at its critical concentration and characterised by an upper critical solution temperature $T_c$: for $T>T_{c}$ the system is in one phase and two phases are obtained whenever $T<T_{c}$. Critical ECF are present in the bulk both above $T_{c}$ in the one-phase region and also below $T_{c}$, in each one of the two phases. In the latter case, capillary ECF at the interface between the two bulk phases are also present \cite{Cicuta:2000cr}. The bulk critical ECF are characterised by the fact that their spatial correlation length $\xi$ and their compressibility $\chi$  diverge when approaching $T_{c}$:

\begin{equation}
\chi=\chi_{0}\epsilon^{-\gamma}\label{eq: Critical compressibility}
\end{equation}

\begin{equation}
\xi=\xi_{0}\epsilon^{-\nu}\label{eq: Critical correlation length}
\end{equation}

where $\epsilon=\left(T-T_{c}\right)/T_{c}$
is the reduced temperature. Here $\chi_{0}$ is a positive constant, $\xi_{0}$ is the bare correlation length and $\gamma$ and $\nu$ are the corresponding critical exponents. The divergence of the compressibility and of the correlation length are also reflected by the behaviour of the structure factor of the concentration fluctuations, which is proportional to the intensity of light scattered by the system. The behaviour of the structure factor for arbitrary values of $q\xi$ is complicated \cite{fisher64,chang79,burstyn83}. However, for small values of the product $q\xi$, the structure factor is well described by the Ornstein and Zernike expression \cite{fisher64}

\begin{equation}
 S_{eq}\left({q}\right)=\frac{k_{B}T}{\rho}\frac{\chi}{1+{q}^{2}\xi^{2}}\label{eq:calmettes}
\end{equation}

where $k_{B}$ is the Boltzmann constant and $\rho$ is the density. According to Eq.\ref{eq:calmettes}, at wave vectors $q\xi>1$ the scattered intensity exhibits a power law behaviour, reflecting the scale invariance of the system at length scale smaller than the correlation length. By contrast, for $q\xi\ll1$, the structure factor becomes q-independent $S\left({q}\right)\propto\chi_{0}\epsilon^{-\gamma}$ and the divergence of the compressibility determines a divergence of the light scattered in the forward direction close to the critical point. All the scattering measurements performed in this work are made in the regime $q\xi\le0.16$ i.e. we expect to observe a q-independent scattering signal from the critical ECF (see also Fig. \ref{fig:deq}b).

The critical exponents in a binary liquid mixture are in the same universality class as the 3D Ising model. The most recent theoretical and computational works \cite{Pelissetto:2002cs,Sengers2009} have fixed the critical exponents to
$\gamma=1.239\pm0.002, \nu=0.630\pm0.002$ and $\eta=0.033\pm0.004$, where $\eta$ is an exponent that accounts for deviations from the ideal Ornstein and Zernike behaviour in Eq. \ref{eq:calmettes}. These exponents are not independent, since the theoretical hyperscaling relation is expected to hold between them:

\begin{equation}
\gamma=\left(2-\eta\right)\nu.\label{eq: gamma nu eta}
\end{equation}

The critical behaviour is also reflected by the dynamics of the concentration fluctuations, which are characterised by a decay rate \cite{sengers1985,onuki02,burstyn83}
\begin{equation}
\Gamma(q)\cong R\frac{k_{B}T}{6\pi\eta_{s}\xi}K\left(q\xi\right)q^{2}\left(1+q^{2}\xi^{2}\right)\label{eq: gamma con kawasaki}
\end{equation}

where $\eta_{s}$ the shear viscosity, $R$ is a universal amplitude close to unity ($R=1.07$ from mode coupling theory and $1.01$ from experiments \cite{sengers1985,onuki02}) and
$K\left(q\xi\right)$ is the Kawasaki scaling function. The Kawasaki function exhibits the two asymptotic behaviours
\begin{equation}
q\xi\ll1\rightarrow K\left(q\xi\right)\cong1,\quad\quad q\xi\gg1\rightarrow K\left(q\xi\right)\cong\frac{3\pi}{8q\xi}.\label{eq: kawasaki}
\end{equation}
For $q\xi\ll1$, the range explored in this work, Eq. \ref{eq: gamma con kawasaki} simplifies to
\begin{equation}
\Gamma(q)\cong R D_{\xi} q^2 
\label{eq: gammaKawasaki}
\end{equation}

where \begin{equation}
  D_{\xi}=\frac{k_{B}T}{6\pi\eta_{s}\left(T \right) \xi\left( T\right) }
  \label{dchitammuo}
\end{equation}
 is the so-called critical diffusion coefficient\cite{hamano86,shmitz95}

The shear viscosity $\eta_{s}$ in Eq.\ref{eq: gammaKawasaki} depends on the temperature of the system and affects the temperature dependence of the diffusion coefficient \cite{berge70}. Very close to the critical temperature, the shear viscosity exhibits a power law dependence on the correlation length \cite{will99,folk2006critical}:

\begin{equation}
\eta_{s}=\bar{\eta}_{s}\left(Q\xi\right)^{z}\label{eq: viscosity power z}
\end{equation}
where $\bar{\eta}_{s}$ is the background viscosity, which may be represented by the Andrade equation $\bar{\eta}_{s}=A\exp\left(B/T\right)$, $Q$
is a system-dependent amplitude and $z$ is a universal exponent,
with a currently accepted theoretical value of $z=0.0679\pm0.0007$ \cite{folk2006critical}. By combining Eq. \ref{eq: viscosity power z} and Eq. \ref{eq: Critical correlation length}
we obtain
\[
\eta_{s}\propto A\exp\left(B/T\right)\left(\frac{T-T_{c}}{T_{c}}\right)^{-y}
\]
where $y=z\nu$. This suggests the opportunity of reformulating the critical behaviour of the diffusion coefficient through an effective exponent $\phi=\nu (z+1)= 0.673\pm0.007$ that takes into account also the temperature dependence of the shear viscosity, so that very close to the critical point one has \cite{shmitz95}:
\begin{equation}
D_{\xi}=D_{0}\left(\frac{T-T_{c}}{T_{c}}\right)^{\phi}\label{eq: diffusion coefficent nu*}
\end{equation}

\subsection*{Non-equilibrium concentration fluctuations (NCF)}   
\label{necf}

The presence of a macroscopic concentration gradient across a binary liquid mixture gives rise to giant non-equilibrium concentration fluctuations \cite{De-Zarate:2006qe}. These fluctuations originate from the coupling of the velocity fluctuations to the concentration gradient. Equilibrium velocity fluctuations determine the local displacement of parcels of fluid, which give rise to small vortices. In the presence of a concentration gradient, the displacement of the fluid parallel to the gradient generates concentration fluctuations, which are not present at equilibrium. In the absence of gravity, these fluctuations exhibit a generic-scale invariance \cite{Grinstein06,brogioli16} reflected in the power law behaviour of the non-equilibrium structure factor that is given by

\begin{equation}
S_{ne}^{g=0}\left(q\right)=\frac{k_{B}T}{\rho}\frac{\nabla c^2}{\nu_s D q^{4}}\label{eq: nogravity}
\end{equation}

 where $\nabla c$ is the amplitude of the macroscopic concentration gradient, $\nu_s$ is the kinematic viscosity and $D$ the mass diffusion coefficient. It is worth stressing that in the presence of a concentration gradient, properties like the density or the diffusion coefficient may vary spatially across the layer, because of their dependence on concentration. This dependence is typically neglected and average values across the layer are usually considered to be representative values for these properties. In practice, when the concentration gradient is small, this is equivalent to consider the value at the center of the cell, where the average concentration remains constant.
 
 The behaviour summarized by Eq. \ref{eq: nogravity} was first predicted theoretically using linearized hydrodynamics by Law and Nieuwoudt \cite{Law89,Nieuwoudt90} in the case of a concentration gradient induced by the Soret effect \cite{deGroot62}. In the absence of gravity, an ultimate limit preventing the divergence of fluctuations at small wave vectors is determined by the finite size of the system \cite{dezarate04,Vailati:2011qf,Cerbino:2015rt,Croccolo:2016mz}. Remarkably, the first experiments on Earth by the group of Jan Sengers \cite{Li94,Li94b} confirmed experimentally the theoretical predictions of Eq. \ref{eq: nogravity}, despite the presence of gravity. The reason is that the effect of gravity becomes apparent in $S_{ne}$ only at very small wave-vectors. In fact, when gravity is considered Eq. \ref{eq: nogravity} becomes

\begin{equation}
S_{ne}\left(q\right)=\frac{k_{B}T}{\rho}\frac{\nabla c\left( \nabla c - \nabla c_{grav} \right)}{\nu_s D }\left[\frac{1}{q_{ro}^{4}+q^{4}}\right]\label{eq: intensit=0000E0 bulk non equilibrium}
\end{equation}

where
\[
\mathbf{\nabla}c_{grav}=\beta\mathbf{g}\chi
\]
is the equilibrium concentration gradient induced by barodiffusion
\cite{Landau} and  \begin{equation}
q_{ro}=\left(\frac{\beta{g}{{\nabla}}c}{\nu_s D}\right)^{\frac{1}{4}}\label{eq: q roll off}
\end{equation}
is a roll-off wave vector determined by the gravity acceleration $g$,  and $\beta$ the solutal expansion coefficient. Eq. \ref{eq: intensit=0000E0 bulk non equilibrium} states that in the presence of a stabilising concentration gradient induced by the Soret effect the buoyancy force determined by gravity quenches concentration fluctuations at small wave-vector. This was first predicted theoretically in Ref. \cite{Segre93} and later observed experimentally in Ref. \cite{Vailati96}. Conversely, in the presence of a destabilising concentration gradient the same gravitational effect gives rise to an amplification of fluctuations, which eventually gives rise to the onset of a convective instability \cite{giavazzi09}.

The presence of a non-equilibrium condition also affects the relaxation of the fluctuations. It has been shown both theoretically \cite{Vailati98} and experimentally \cite{Croccolo07,Giavazzi:2016fr} that for $q>q_{ro}$ the relaxation of fluctuations occurs by diffusion, while at smaller wave vectors the relaxation is dominated by the drag effect determined by the gravity force, so that the decay rate of the fluctuations is given by

\begin{equation}
\Gamma(q)=Dq^{2} \left[1+\left( \frac{q_{ro}}{q} \right)^4  \right].
\label{eq: gamma con Kawasaki nel nostro caso}
\end{equation}

In principle, at even smaller wave vectors the finite size of the sample restores a diffusive relaxation with an effective diffusion coefficient, as recently shown in Ref. \cite{Giraudet15}. This regime is however not accessible to our experiments.

Theoretical models and experiments relied at first on stationary macroscopic non-equilibrium states induced by generating a concentration gradient by means of the Soret effect. Further experimental investigation showed that non-equilibrium concentration fluctuations can be induced generically by the presence of a macroscopic concentration gradient, independently of the way the gradient is created \cite{Vailati:1997kq,Brogioli00}. The first evidence of this fact was achieved by taking advantage of the unique features of a near critical binary liquid mixture, which allow inducing a strong concentration gradient free of disturbances simply by changing the temperature of the sample. The sample is initially let separate into two phases at $T<T_{c}$. The temperature is then suddenly raised above $T_{c}$, so that the two phases become miscible. The two phases are initially separated by a sharp interface and gradually mix through a diffusion process until the sample attains the critical concentration $c_{c}$. During the early stages of the diffusion process, large amplitude capillary fluctuations develop at the interface \cite{cicuta01}.
At later stages, the strong concentration gradient that drives the diffusion process gives rise to large-amplitude bulk concentration fluctuations. \\ Using a critical fluid is thus an interesting trick to trigger NCF. However, we note that special care has to be used in the interpretation of the non-equilibrium data, because the thermophysical properties of the mixture become strongly dependent on temperature and concentration close to a critical consolute point.

The theoretical description of these time dependent fluctuations required to generalize the steady state models based on linearized hydrodynamics to the time dependent case. The use of an adiabatic approximation allows to separate the equations for the evolution of the macroscopic state from the equations describing the dynamics of the fluctuations. Under this approximation the static structure factor is given by \cite{Vailati98}

\begin{equation}
S\left(q\right)=S_{eq}\left[1+\left(\frac{\mathbf{\nabla}c}{\nabla c_{grav}}-1\right)\frac{1}{1+\left(q/q_{ro}\right)^{4}}\right]\label{eq: non equilibrium structure factor vailati}
\end{equation}
where now  $S_{eq}$, $\nabla c_{grav}$,$\nabla c$ and $q_{ro}$ depend upon the macroscopic state and are thus functions of $\left(z,t\right)$. 

In the case of near critical binary mixture, a peculiar non-equilibrium condition can be attained by letting the sample separate into two phases of concentration $c_{i}^{+}$ (upper phase) and $c_{i}^{-}$ (lower phase)  at a temperature $T_i<T_{c}$. The temperature is then suddenly raised to a value $T_f$, with $T_i<T_f<T_{c}$. In the final state the two phases have concentrations $c_{f}^{+}$  and $c_{f}^{-}$, where $c_{i}^{+}< c_{f}^{+}<c_{f}^{-}<c_{i}^{-}$. Both in the initial and final states the two phases are separated by a sharp interface. The evolution from one state to the other involves a diffusive mass transfer between the two phases, which determines bulk non-equilibrium fluctuations in the two phases \cite{Cicuta:2000cr}. 
We note that, if the mixture is prepared at its critical concentration, because of the lever rule discussed in the next section, the volumes of the two phases are equal and this implies that the position of the interface remains fixed during diffusion.

The features of these non-equilibrium fluctuations are affected by the time evolution of the macroscopic state as a function of both the vertical coordinate $z$ and the time $t_{d}$ elapsed from the beginning of the diffusion process. An analytical series expansion for $c\left(z,t_{d}\right)$ during a free diffusion experiment in which two initially separated phases are free to remix can be obtained by solving the diffusion equation with initial and final conditions  $c_{i}\left(z\right)$ and $c_{f}\left(z\right)$ determined by the Heaviside step functions \cite{crank}:

\begin{equation}
c_{i,f}\left(z\right)=
\begin{cases}
c_{i,f}^{-}, & 0<z<h\\
c_{i,f}^{+}, & h<z<a
\end{cases}
\end{equation}

where $a$ is the sample thickness and $h=a/2$ is the position of the interface between the two phases.

The corresponding time evolution of the concentration profile for a mixture prepared at its critical concentration is (Fig. \ref{fig:dnoneq}b))
\begin{multline}
c\left(\zeta,t_{d}\right)=
c_{f}\left(\zeta\right)+\\
+\frac{2}{\pi}\left(c_{f}^{+}-c_{f}^{-}\right)\sum_{j}^{\infty}\frac{\sin\left(\frac{j\pi h}{a}\right)\cos\left(j\pi\zeta\right)}{j}e^{-\frac{Dj^{2}\pi^{2}}{a^{2}}t_{d}}\label{eq: c(zeta,t) step diffusion}
\end{multline}
where $\zeta=z/a$.

\section*{Materials and methods}
\label{mem}
\subsection*{The sample: preparation and properties} 
\label{sample}

\begin{figure}[hbt]
  \includegraphics[width=0.95\columnwidth]{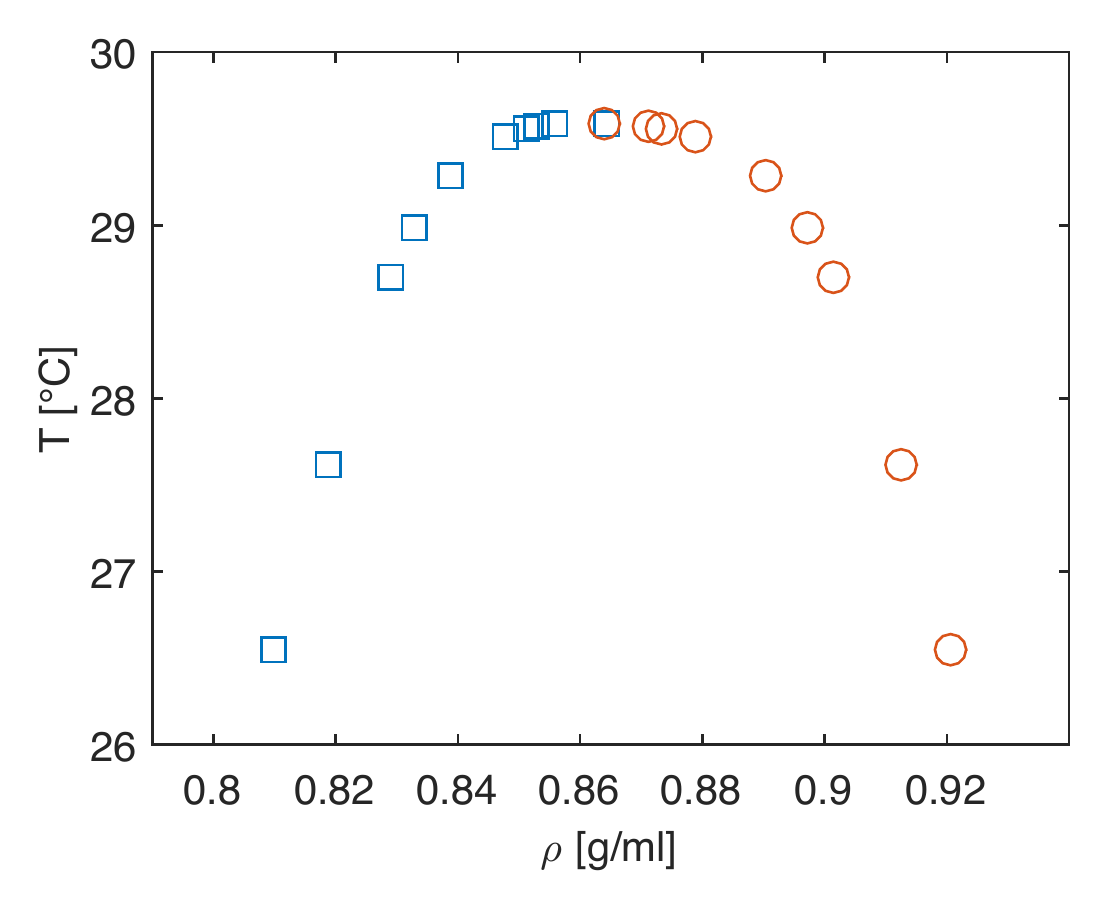}
  \caption{Experimental coexistence curve of the mixture aniline-cyclohexane. Data from Ref. \cite{atack53}}
  \label{fig: Coexistence-curve}
\end{figure}

The sample is the  binary mixture aniline and cyclohexane, prepared at its critical aniline concentration  $c_{c}=0.47\,w/w$. This critical fluid presents a coexistence curve (Fig. \ref{fig: Coexistence-curve})
with a critical temperature $T_{c}\approx 30\degree$C that divides an upper homogeneous phase region and a lower two-phase region, in which an aniline-rich phase and a cyclohexane-rich phase are separated by a sharp interface \cite{atack53}.
The thermophysical properties of this binary mixture have been determined accurately by several research groups (table \ref{tab:1}). 

\begin{table*}
\caption{Typical thermophysical properties of the binary liquid mixture aniline/cyclohexane prepared at its critical concentration $c_c=0.47$ aniline weight fraction.}
\label{tab:1}       
\begin{tabular}{lll}
\hline\noalign{\smallskip}
Property & Expression & parameters  \\
\noalign{\smallskip}\hline\noalign{\smallskip}

correlation length & $\xi=\xi_0 \epsilon^{-\nu}$ & $\xi_0=0.22$ nm, $\nu=0.63$ \cite{calmettes72} \\

compressibility & $\chi=A_V \left( T-T_C\right)^{-\gamma}$  & $A_V=7.73\times 10^{-7}$ (cgs units), $\gamma=1.2$ \cite{giglio1975b}\\

diffusion coefficient & $ D=D_0 \epsilon^{\phi}$ & $D_0=1.42 \times 10^{-5}$ cm $^2$/s,  $ \phi=0.73 $ \cite{giglio1975}\\

mass density & $\rho=\frac{\rho_A \rho_C}{c \rho_A +\left( 1-c\right) \rho_C}$ & $\rho_A=1.024$ g/cm$^3$, $\rho_C=0.78$ g/cm$^3$ \cite{strandgaard_databook}\\

specific heat at constant pressure & $c_p=X c_{p,A}+ \left( 1-X\right)c_{p,C}$  &  $c_{p,A}=2.08 J/\left( g K\right),  c_{p,C}=1.81$ J/(g K), $X$: A mol. frac. \cite{strandgaard_databook}\\

expansion coefficient & $\alpha=c \alpha_A+(1-c)\alpha_C$ & $\alpha_A=0.85\times 10^{-3}$ K$^{-1}$, $\alpha_C=1.2\times 10^{-3}$ K$^{-1}$  \\

solutal expansion coefficient & $\beta=\frac{1}{\rho}\left( \frac{\partial \rho}{\partial c}\right)_{p,T}$ & 0.273 at $c=c_c$  \\

kinematic viscosity & $\nu_s=\frac{\eta_{s}}{\rho}$ & $\nu_s=1.35 \times 10^{-2}$ cm$^2$/s, $\eta_{s}=1.2 \times 10^{-2}$ g/(cm s) at $T=40\degree$C\\

\noalign{\smallskip}\hline
\end{tabular}
\end{table*}

It is known that volumes and concentrations of the coexisting phases are not independent on each other. They instead satisfy a specific relation, called \textit{lever rule}, that originates from conservation of the total mass of the system. If $c_{0}$ is the concentration of the sample, $\phi^{+}$ and $c^{+}$ are the volume and the concentration of the upper phase, respectively, and $\phi^{-}$ and $c^{-}$ the volume and the concentration of the lower phase, respectively, we have

\begin{equation}
\phi^{+}=\frac{c_{0}-c^{-}}{c^{+}-c^{-}},\ \ \ \ \ \phi^{-}=\frac{c^{+}-c_{0}}{c^{+}-c^{-}}\label{eq: lever rule}
\end{equation}

and 
\begin{equation}
\frac{\phi^{+}}{\phi^{-}}=\frac{c_{0}-c^{-}}{c^{+}-c_{0}}.\label{eq: lever rule ratio}
\end{equation}

The coexistence curve of aniline and cyclohexane is symmetric for
temperature close to $T_{c}$, i.e. for all the temperature range
shown in Fig. \ref{fig: Coexistence-curve}. As a consequence, for
$T$ close to $T_{c}$ and for $c_{0}=c_{c}$ 
\[
c_{0}=c_{c}=\frac{c^{+}+c^{-}}{2}
\]

and 
\[
\frac{\phi_{+}}{\phi_{-}}=1
\]

\textit{i.e.} the volumes of the two phases are equal if the sample is prepared at the critical concentration. This is not the case for $T\ll T_{c}$ for which the asymmetry of the coexistence curve emerges as a consequence of the different densities of aniline and cyclohexane.

\subsection*{Sample cell and temperature control} 
\label{cell}

\begin{figure}[hbt]
  \includegraphics[width=0.95\columnwidth]{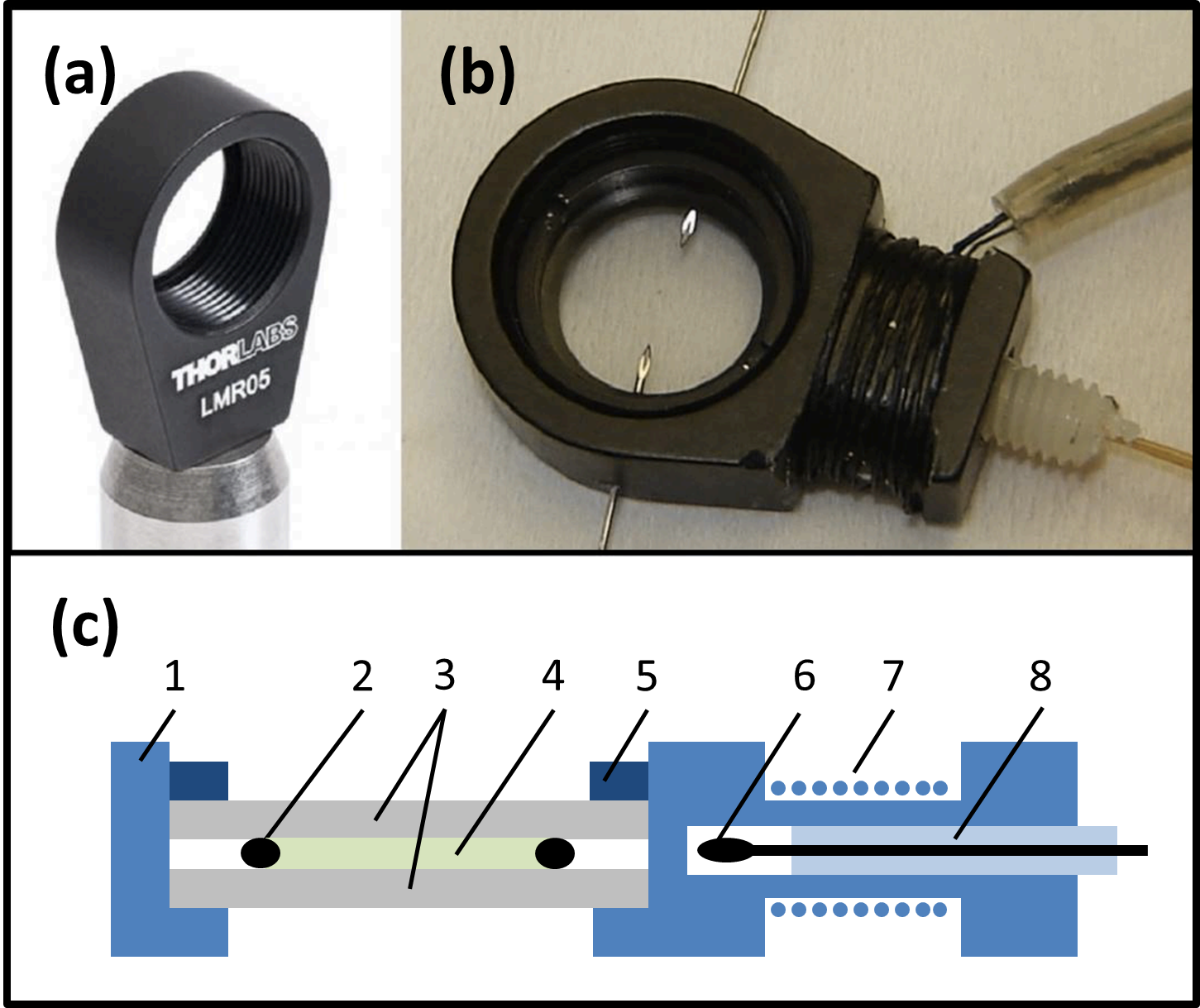}
  \caption{a) Commercial aluminium lens holder used in the construction of the sample cell. b) Sample cell. The two hypodermic needles, visibile in upper and in the lower part of the image, are removed after the filling of the cell. c) Schematic cross-sectional view of the assembled sample cell: 1-lens holder, 2-O-ring, 3-sapphire windows, 4-sample, 5-retaining ring, 6-thermistor, 7-heating coil, 8-hollow nylon screw.}
\label{fig: CELL}
\end{figure}

The sample cell needs to meet several requirements: \textit{i)} provide optical access to the sample, \textit{ii)} have a small overall thickness for mounting the cell under the microscope with sufficient clearance for positioning the objective to image the liquid slab, \textit{iii)} be chemically resistant to the critical  binary mixture  of aniline and cyclohexane \textit{iv)} maintain the  sample at a uniform temperature.  To achieve simultaneously a good thermalization, optical clearance and chemical resistance to the aniline/cyclohexane mixture the sample is sandwiched between two cylindrical sapphire windows of diameter $12.5$ mm and thickness $0.5$ mm, the lateral confinement being provided by a Viton O-ring of thickness $1.7$ mm.  The outer part of the cell is made with a commercial aluminium lens holder,  suitably modified for hosting the sapphire windows, a resistive heating coil and a thermistor (Fig. \ref{fig: CELL}).

The filling of the cell is achieved by means of two hypodermic needles (diameter $0.4$ mm) laterally inserted into the Viton O-ring. 

The thermalization  of the cell is achieved by a nickel-chrome resistive wire coil wrapped inside a wide groove machined in the stem of the lens holder. The temperature of the cell is read by a thermistor placed inside a hole in the stem of the lens holder. The thermistor is connected to a commercial temperature controller (LFI-3751, Wavelength Electronics, USA), which drives the nickel-chrome wire with a current whose intensity is controlled by a Proportional Integral Derivative servo. The overall stability of the temperature achieved in this way is of the order of $10$ mK over a few hours.

\subsection*{Differential Dynamic Microscopy}

Scattering information relative to the concentration fluctuations
was obtained by Differential Dynamic Microscopy (DDM) \cite{Cerbino:2008if,Giavazzi:2009xd}.
DDM is a novel light scattering method \cite{scheffold2007new} that
relies on the fact that time-lapse movies made of images collected
within the sample or in its close proximity \cite{Cerbino:2009ta}
do contain the holograpic footprint of the sample concentration distribution,
as well as the information about the temporal correlation properties
of concentration fluctuations. Reciprocal space information can be
extracted from the $N$ intensity maps of the video $i_{n}(\mathbf{x})=i(\mathbf{x},n\delta t)$
acquired at times $n\delta t$ ($n=1,...,N$) by calculating their
spatial Fourier transform $I_{n}(\mathbf{q})=\int e^{-j\mathbf{q}\cdot\mathbf{x}}i_{n}(\mathbf{x})d\mathbf{x}$.
Here $j$ is the imaginary unit, $\delta t$ is the inverse frame rate of the
video and $\mathbf{x}=(x,y)$ and $\mathbf{q}=(q_{x},q_{y})$ are the
real space and reciprocal space coordinates, respectively. In DDM
experiments, the spatial and temporal correlation of concentration
fluctuations is extracted by using a differential algorithm,
first proposed in Ref. \cite{Croccolo:2006kl}. The key quantity
of the DDM analysis is the image structure function defined for each
time delay $\Delta t=m\delta t$ as
\begin{equation}
d(q,\Delta t)=\left\langle \left\langle \left|I_{n+m}(\mathbf{q})-I_{n}(\mathbf{q})\right|^{2}\right\rangle _{n}\right\rangle _{q=\sqrt{q_{x}^{2}+q_{y}^{2}}},\label{eq:imstrfunc}
\end{equation}
where the average $\text{\ensuremath{\left\langle ...\right\rangle _{n}}}$
is made over image pairs that are separated by the same time delay
over the entire stack. Also, the azimuthal average $\text{\ensuremath{\left\langle ...\right\rangle _{q=\sqrt{q_{x}^{2}+q_{y}^{2}}}}}$
is typically performed when, as in the present case, the structure
and the dynamics of the sample are isotropic and spatially homogeneous
in the image plane. It has been shown that $d(q,\Delta t)$ is directly
connected to the normalized intermediate scattering function $f(q,\Delta t)$
by the relation
\begin{equation}
d(q,\Delta t)=A(q)\left[1-f(q,\Delta t)\right]+B(q),\label{eq:strint}
\end{equation}
where $B(q)$ is a background contribution due to the noise of the
detection chain and $A(q)=I(q)T(q)$ is an amplitude that depends
on the scattering intensity $I(q)$ and on a transfer function $T(q)$
describing the microscope \cite{Giavazzi:2014it}. In principle,
$T(q)$ can be determined by calibration of the microscope with a
suitable sample, as done for instance in Ref. \cite{Giavazzi:2016fr}.
Another route, which is the one employed here, is to divide out curves
with the same $T(q)$ and different scattering intensity $I(q)$ to
obtain a relative intensity measurement. Further details will be provided
later on in the text. 

The normalized intermediate scattering function $f(q,\Delta t)$ describes
the dynamic correlation properties of the sample and can be calculated
theoretically for a variety of cases \cite{Berne:1976yu}. In many cases of interest, concentration fluctuations relax monoexponentially \textit{i.e.}
\begin{equation}
  f(q,\Delta t)=e^{-\Gamma(q)\Delta t}
  \label{modelfit}
\end{equation}
where $\Gamma(q)$ is the correlation rate of the fluctuations. For instance, relaxation by diffusion is given by Eq. \ref{modelfit} with $\Gamma(q)=D q^{2}$, where $D$ is the diffusion coefficient.

One important feature of DDM compared to other Digital Fourier Microscopy techniques \cite{Giavazzi:2014it} is that the use of a partially coherent light source reduces the depth of field of the imaging system compared to using a laser light source \cite{Giavazzi:2009xd,Giavazzi:2014it}. In practice, this means that in a stratified system one can select a thin region along the microscope optical axis and obtain scattering information only in this region, whose thickness is set by the numerical aperture of the microscope condenser. This was recently done for instance in Ref. \cite{buzzaccaro2013ghost} where this idea was used to perform three-dimensional reconstructions of flows, also in the absence of tracers, by using a commercial microscope. Here, we make use of the reduced depth of field of our microscope to select a suitable region of the sample along the vertical direction, which in our case is particularly important for $T<T_{c}$, when the system is in two phases.

In this work, we analyze with DDM several image stacks captured at $200$ fps with a standard inverted bright-field microscope (Nikon Ti-U), equipped with a CMOS camera (Hamamatsu ORCA-Flash4.0, pixel size $6.5$ $\mu m$). Each stack comprises $50000$ images ($128\times128$ pixels) and corresponds to an acquisition made at a different temperature for the experiments performed in the one phase region, or at a different time $t_{d}$ from the beginning of diffusion for the experiment performed in the two-phase region. For the former experiments, we used 2x2 binning and an objective magnification equal to 20x, whereas for the latter we used a 10x objective without binning. The use of a large number of images turned out to be beneficial to achieve a reliable statistical characterisation of the ECF far away from the critical point, where the scattering signal $A$ was about one order of magnitude smaller than the camera noise $B$ (see Eq. \ref{eq:strint}).

\section*{Results}

\subsection*{Concentration fluctuations in the one-phase region}

\begin{figure*}[hbt]
\begin{centering}
\includegraphics[width=1\textwidth]{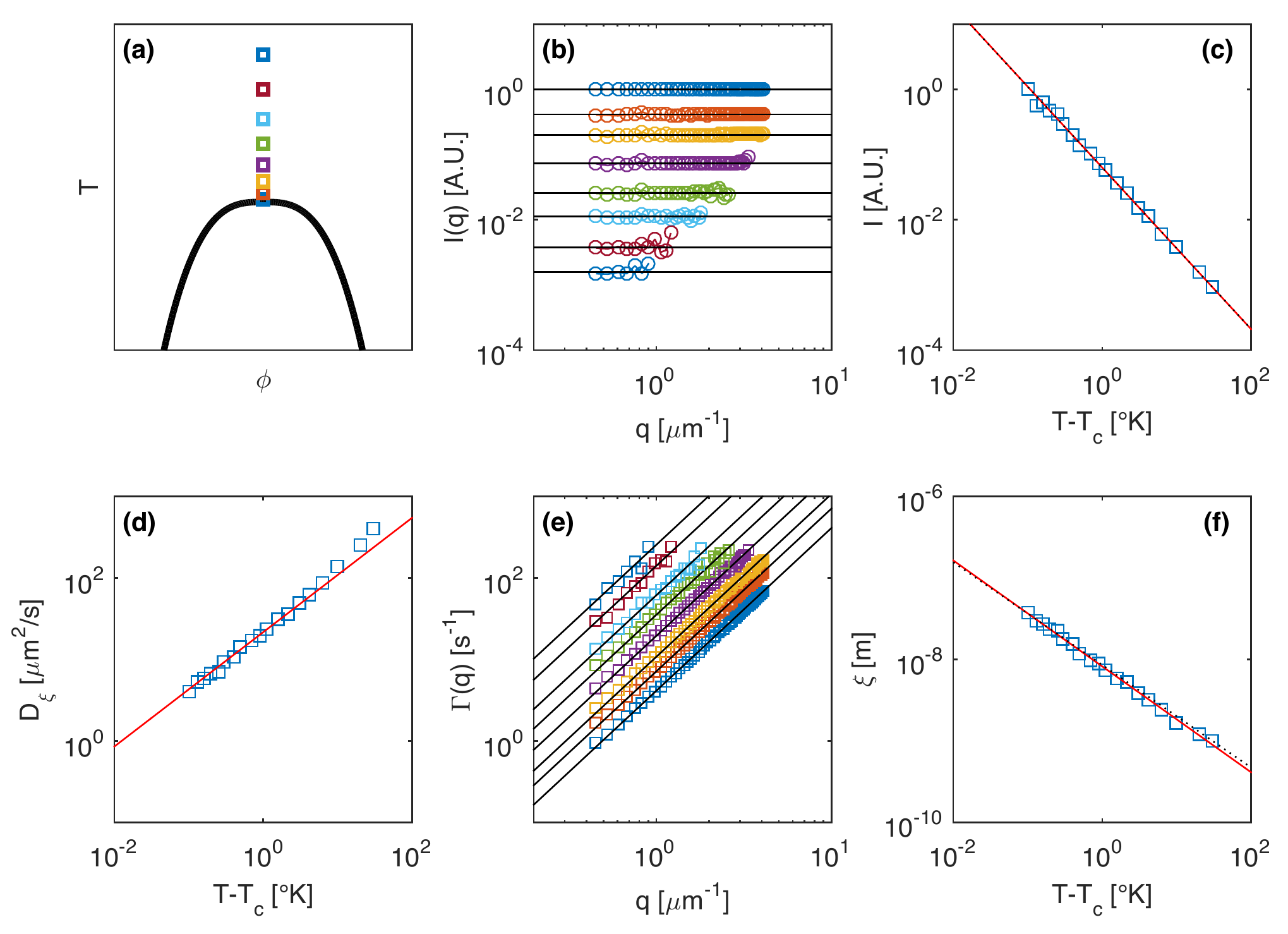}
\par\end{centering}
\caption{ a) Pictorial representation of the phase diagram of the binary mixture used in this work. Each square corresponds to an experiment. All the experiments are performed at the critical concentration, at different temperatures above the critical temperature $T_{c}$. For clarity we show only selected temperatures. b) Scattering intensity $I(q)$ of the critical fluctuations at various temperatures above the critical temperature,as in panel a. c) Critical behaviour of the scattered intensity $I$ as a function of the temperature difference $T-T_{c}$. Continuous and dotted lines are best fit of the data to a power-law curve with fixed and free exponents, respectively (see text for details). d) Diffusion coefficient  $D_{\xi}$ obtained from the study of the dynamics of the fluctuations. The continuous line is a best fit with Eq. \ref{eq: diffusion coefficent nu*} . e) $q$-dependent relaxation rates for the same temperatures of panel a. Continuous lines are best fits with a function $D_{\xi}q^{2}$. f) Critical behaviour of the correlation length $\xi$, calculated from the data according to Eq. \ref{eq:xi}. Continuous and dotted lines are best fit of the data to a power-law curve with fixed and free exponents, respectively (see text for details).}\label{fig:deq}
\end{figure*}

In a first sets of experiments, we probed the equilibrium concentration fluctuations (ECF) exhibited when the sample is in one phase i.e. when $T>T_{c}$. We used DDM to obtain scattering information in the temperature range $0.1^{o}C\,\leq T-T_{c}\leq30^{o}C$. More precisely, we performed measurements at $T-T_{c}=$30.1, 20.1, 10.1, 6.1, 4.1, 3.1, 2.1, 1.6, 1.1, 0.9, 0.7, 0.5, 0.4, 0.3, 0.25, 0.2, 0.16, 0.13, 0.1 $^{o}$C, some representative temperatures being schematically reported in Fig. \ref{fig:deq}a. Each stack was processed with the DDM algorithm described above and the image structure functions obtained where fitted by using Eqs. \ref{eq:strint} and \ref{modelfit} that enabled us extracting for each temperature, a $q$-dependent amplitude $A(q)$ and a $q$-dependent relaxation rate $\Gamma(q)$, as well as an estimate of the camera noise $B(q)$. 

In Fig. \ref{fig:deq}b we show the scattered intensity $I(q)$ for selected values of $T-T_{c}$, which is obtained, up to a multiplicative constant, as the ratio $I(q)=\frac{A(q)}{A_{0}(q)}$, where $A_{0}(q)$ is a reference amplitude obtained in a condition where $I(q)=const=I_{0}$. In this work we have chosen $A_{0}(q)$ as the amplitude associated with the smaller investigated value of $T-T_{c}=0.1$ $^{o}$C, condition in which the amplitude of the fluctuation is the largest, allowing an accurate determination of the microscope transfer function. We note that, even so close to $T_{c}$, the correlation length $\xi$ is expected to be in the nanometer range and this ensures that $I(q)$ is substantially $q$- independent in the investigated $q$-range. The effectiveness of this assumption can be checked in Fig. \ref{fig:deq}b, where we observe that the scattering intensity is found to be constant for all the investigated temperatures. This allows a straightforward determination of the forward scattering intensity $I$, obtained as an average of $I(q)$ over the whole investigated $q$-range.

The temperature dependence of the forward scattering intensity $I$ exhibits a very neat power law dependence on $T-T_{c}$, over the quite impressive range \textbf{$[0.1,30]$} $^{o}$C (Fig. \ref{fig:deq}c). According to Eq. \ref{eq:calmettes}, the scaling behaviour of $I$ is expected to be determined by same the critical exponent $\gamma$ describing the divergence of $\chi$ when approaching the critical point. A fit of the experimental data to a simple power law curve $I=a\,(T-T_{c})^{-\gamma_{exp}}$ (dotted black line in Fig. \ref{fig:deq}c) provides the estimate $\gamma_{exp}=1.24\pm0.01$. This value is in excellent agreement with the value $\gamma=1.239\pm0.002$ theoretically predicted for the 3D Ising model universality class. We also show in Fig. \ref{fig:deq}c the curve $I=a'\,(T-T_{c})^{-\gamma}$ as a red line. The two best fitting curves, with free and fixed exponent, are barely distinguishable.

In Fig. \ref{fig:deq}e we report the relaxation rate of the fluctuations $\Gamma(q)$ for selected values of $T-T_{c}$. Continuous lines represent best fitting curves to the data with Eq. \ref{eq: gammaKawasaki}, where the value of the universal amplitude $R$ was set to its mode-coupling theory estimate of 1.07. The behaviour of the critical diffusion coefficient $D_{\xi}$ as a function of temperature is reported in \ref{fig:deq}d, where the critical slowing down of the dynamics as the critical temperature is approached becomes very evident. The continuous line is the best fit of the data to Eq. \ref{eq: diffusion coefficent nu*}, from which the value $\phi_{exp}=0.70\pm 0.02$ is obtained, in fair agreement with the best estimate $\phi=0.673\pm 0.007$ . We note that the data are well described by a power law only very close to the critical temperature, while a systematic deviation is observed for  $T-T_{c} > $2$^{o}$C, in agreement with previously reported measurements \cite{giglio1975}. This deviation is not surprising, since Eq. \ref {eq: diffusion coefficent nu*} is obtained by neglecting the exponential temperature dependence of the background viscosity $\bar{\eta}_{s}$, as well as the explicit temperature dependence of $D_{\xi}$ in the numerator of the rhs of Eq. \ref{dchitammuo}.  The above-reported estimate for the critical exponent $\phi_{exp}$ is thus obtained by a fit restricted to the data collected in the interval $0.1<T-T_{c}<1.6$ $^{o}$C.

In order to decouple the contributions to the critical diffusion coefficient from the shear viscosity $\eta_{s}$ and the correlation length $\xi$ of the concentration fluctuations, it can be worth to recast this result in terms of the correlation length, by making use of Eq. \ref{dchitammuo} and using for $\eta_{s}(T)$ values obtained from the literature \cite{arcovito69}. By inverting Eq. \ref{dchitammuo} we obtain

\begin{equation}
\xi\left(T\right)=\frac{k_{B}T}{6\pi\eta_{s}\left(T\right)D_{\xi}(T)}\label{eq:xi}
\end{equation}

The scaling of $\xi$ as a function of the reduced temperature is reported in Fig. \ref{fig:deq}f, together with the best fitting curves to the data obtained with the function $\xi=\xi_{0}\left(\frac{T-T_{c}}{T_{c}}\right)^{-\nu_{exp}}$. The black dotted curve is obtained by using both the critical exponent $\nu_{exp}$ and the bare correlation length $\xi_{0}$ as free fitting parameters, leading to $\xi_{0}=20\pm3\,\textrm{nm}$ and $\nu_{exp}=0.65\pm0.02$. The red line is obtained by imposing for the critical exponent the 3D Ising value $\nu_{exp}=\nu=0.630$. The estimate for the bare correlation length obtained in this case is $\xi'_{0}=23.5\pm0.5\,\mbox{nm}$. Although with a larger relative uncertainty, even in this case the experimentally measured value of the critical exponent is in agreement with the theoretical prediction.  Also, we note that in this case no systematic deviation from a simple power law behavior can be outlined over the whole investigated temperature range, indicating that Eq. \ref{eq: Critical correlation length} holds in a wider neighbourhood of $T_{c}$.

We conclude this part by noting that quite remarkably the good overall accuracy of our experimental determination of the critical exponents, enables us providing an estimate of the elusive critical correlation-correction exponent $\eta$, which has been considered a severe test bench for all the experimental methods for the study of critical phenomena. This is made possible by exploiting the hyperscaling relation (Eq. \ref{eq: gamma nu eta}) that gives $\eta=2-\gamma/\nu=0.09\pm 0.06$. Although affected by a large error, this value is not compatible with 0 and it is agreement with the theoretical prediction $0.033\pm 0.004$. Incidentally, this value is also very close to the value $0.08\pm 0.01$ reported in the literature for the same system \cite{Sengers2009}. The error on the determination of $\eta$ can be in principle reduced by reducing the error on the determination of $\nu$, for instance by acquiring longer movies or by repeating the experiment several times.

\subsection*{Concentration fluctuations in the two-phase region}

\begin{figure*}[hbt]
\begin{centering}
\includegraphics[width=1\textwidth]{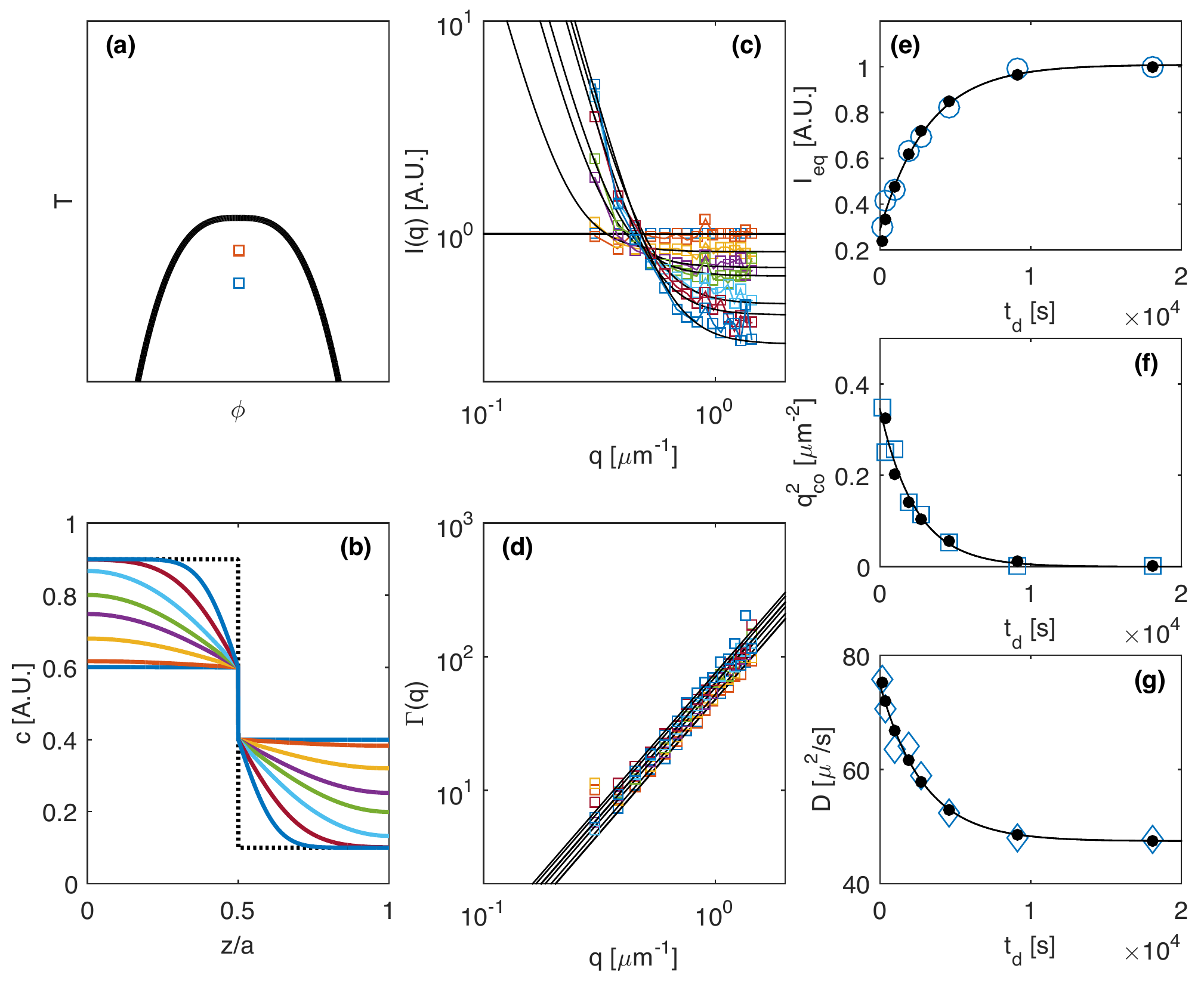}
\par\end{centering}
\caption{a) Pictorial representation of the phase diagram of the binary mixture used in this work. The two squares correspond to the temperatures $T_{i}$ and $T_{f}>T_{i}$ used for triggering the NCF. b) Concentration profiles at $t_{d}=1,5,15,30,45,75,150,300$ min from the beginning of the diffusion process calculated from Eq. \ref{eq: c(zeta,t) step diffusion}. c) Scattering intensity $I(q)$ associated with the concentration fluctuations, measured at the same times indicated in panel b. All the curves exhibit an excess of scattering at small $q$, due to the NCF, and plateau for large $q$ corresponding to the signal originated by ECF. The continuous curves are best fitting curves obtained with Eq. \ref{eq:CROSS}. d) Correlation rate $\Gamma(q)$ measured for the same times indicated in panel b. The continuous curves are best fitting curves obtained with a simple diffusive scaling $\Gamma(q)=Dq^2$. e) Empty symbols: equilibrium scattering intensity $I_{eq}$ measured at various times during diffusion; black dots: scaled average concentration in one of the two phases at the same times calculated from Eq. \ref{eq: c(zeta,t) step diffusion}; continuous line: best fit to the data with an exponential model. f) Empty symbols: squared cross-over wave vector $q^2_{co}$ measured at various times during diffusion; black dots: concentration gradient at the interface between the two phases at the same times calculated from Eq. \ref{eq: c(zeta,t) step diffusion}; continuous line: best exponential fit to the data. g) Empty symbols: diffusion coefficient $D$ measured at various times during diffusion; black dots: scaled average concentration in one of the two phases at the same times calculated from Eq. \ref{eq: c(zeta,t) step diffusion}; continuous line: best exponential fit to the data.}\label{fig:dnoneq}
\end{figure*}

Before starting this experiment, we left the sample equilibrate for at least $24$ hours at a temperature $T_{i}=T_{c}-2.4$ $^{o}$C. In this condition the sample is separated in two phases (of concentrations respectively $c^{-}$ and $c^{+}$), forming two horizontal layers laying one on top of the other, with the denser (aniline-richer) phase lying at the bottom of the cell. We note that, at variance with the previous experiment, the sample is prepared close to but not exactly at $c_{c}$ (concentration $c_{0}$ was about 2$\%$ larger than $c_{c}$). However, it is safe to assume that each one the two phases occupies approximately $1/2$ of the total volume of the mixture and that the position of the interface separating the two layers remains substantially fixed at the middle height of the cell, as far as $T_{i}<T_{c}$.
At the time $t_{d}=0$ the temperature is suddenly raised from $T_{i}=T_{c}-2.4$ $^{o}$C to $T_{f}=T_{c}-1.1$ $^{o}$C (Fig. \ref{fig:dnoneq}a). As described above, the presence of a diffusive mass flow that aims at restoring equilibrium triggers the onset of NCF. We have performed a systematic study of such fluctuations by following in time the evolution of the system during the diffusion process triggered by the change of temperature. This is done by acquiring several stacks of microscope images, each stack being representative of the system at $t_{d}=1,5,15,30,45,75,150,300$ min from the beginning of the diffusion process. We stress once again that the main contribution to the scattering signal from the NCF comes from the center of the cell where the excess concentration gradient is larger. In order to enhance the contrast of the NCF scattering signal we imaged a plane slightly below the interface between the two phases. 

A DDM analysis was performed on each stack. The obtained image structure functions were fitted by using Eqs \ref{eq:strint} and \ref{modelfit}. The fitting process enabled us extracting the amplitude $A(q)$, the relaxation rate $\Gamma(q)$, and the camera noise $B(q)$, at different time $t_{d}$ after the beginning of the diffusion process. In close analogy with the procedure discussed in the previous section, the scattering intensity $I(q)$ is obtained as $A(q)/A_{0}(q)$. In the present case we used as a reference amplitude $A_{0}(q)$ the one measured in the latest stages of the process, when an equilibrium condition is recovered.

Since this experiment was performed to investigate the interplay between ECF and NCF in a binary mixture, the choice of the accessible wave-vector range was dictated by the need of accessing the high-q region, where the scattering amplitude of the NCF becomes smaller than the one of ECF. The rolloff wave-vector $q_{ro}$ lies outside this range, which implies that gravity does not affect the amplitude and the lifetime of the fluctuations. As a results, the structure factor (Eq. \ref{eq: non equilibrium structure factor vailati}) takes a simpler form. By introducing the crossover wave-vector \cite{Giavazzi:2016fr}
\begin{equation}
q_{co}=\sqrt[4]{\frac{\left(\nabla c\right)^{2}}{\nu_s D\chi}}.\label{eq:crossover}
\end{equation}

the corresponding scattering intensity can be written as
\begin{equation}
I(q)=I_{eq}\left[1+(q_{co}/q)^{4}\right],\label{eq:CROSS}
\end{equation}
which we used to fit our scattering data. Similarly, in our wave-vector range, Eq. \ref{eq: gamma con Kawasaki nel nostro caso} for the relaxation rate of the concentration fluctuations simplifies to the simple diffusive  relaxation rate $\Gamma(q)=Dq^2$. 

In Fig. \ref{fig:dnoneq}c we show the scattering intensity $I\left(q\right)$ at different times after the beginning of the diffusion process, together with the best fittings curves with Eq. \ref{eq:CROSS}. Two features of $I(q)$ are immediately apparent. On one hand, the amplitude of the low-$q$ divergent $q^{-4}$ portion progressively decreases in time and $q_{co}$ becomes smaller, as expected from Eq. \ref{eq:crossover}. On the other hand, the high-$q$ plateau, corresponding to the equilibrium contribution, progressively shifts toward larger values, until a saturation value is reached. These two trends can be clearly appreciated also in panels e) and f) where the time evolution of the equilibrium scattering intensity $I_{eq}$ and of the squared cross-over wave-vector $q_{co}^{2}$ are reported, respectively.

$I_{eq}$ accounts for the equilibrium fluctuation occurring in the bulk, and according to Eq. \ref{eq:calmettes} it follows the compressibility $\chi$ of the mixture, which increases when approaching the critical point (also from below $T_{c}$ ). Interestingly, we find that the time evolution of $I_{eq}$ substantially coincides with the time evolution of the average concentration $\bar{c}(t)=\int_{0}^{h}c(z,t)dz$ in one of the two phases (say, the one with the higher concentration). $\bar{c}(t)$ can be estimated from Eq. \ref{eq: c(zeta,t) step diffusion} with $a=1.4$ mm and  by assuming $D=70$ $\mu m^{2}/s$, a representative value obtained by the study of the dynamics of the fluctuations (see panel g)). $\bar{c}(t)$ is  represented as black dots in Fig. \ref{fig:dnoneq}e).  Since the initial and final concentrations are not known, $\bar{c}(t)$ can be determined only up to an additive constant and a scaling factor, which has been adjusted in order to make easier the comparison with $I_{eq}$ in Fig. \ref{fig:dnoneq}e).

According to Eq. \ref{eq:crossover}, the behaviour of $q_{co}^{2}$ as a function of time is mainly determined by the evolution of the concentration gradient, since all the other quantities are expected to have a milder time dependence. This prediction is confirmed by the excellent agreement between the experimentally determined time evolution of $q_{co}^{2}$ and the relaxation of the concentration gradient at the center of the cell (black dots in Fig. \ref{fig:dnoneq}). The estimate of the gradient is obtained again by using Eq. \ref{eq: c(zeta,t) step diffusion} with $D=70$ $\mu m^{2}/s$.

In Fig. \ref{fig:dnoneq}d, we show the $q$-dependent relaxation rate $\Gamma(q)$ obtained for different sampling time $t_{d}$. While for a given $t_{d}$, $\Gamma(q)$ is always well described by a simple diffusive relaxation $\Gamma(q)=Dq^{2}$, the estimated diffusion coefficient shows a marked time dependence (see Fig. \ref{fig:dnoneq}g). As for $I_{eq},$ the change of $D$ over time can be explained by the progressive change in the composition of the sample due to diffusion, as it closely mirrors the evolution of the average concentration $\bar{c}(t)$ in one of the two phases (black dots).

All the quantities in panels e,f,g of Fig \ref{fig:dnoneq} exhibit an exponential relaxation with estimated time constants  $\tau_{I_{eq}}=(3.1\pm0.4) \, 10^{3}$ s,  $\tau_{q_{co}^{2}}=(2.4\pm0.5) \, 10^{3}$ s and $\tau_{D}=(2.9\pm0.6) \, 10^{3}$ s, respectively. All these values  corresponds well to the characteristic relaxation time of the concentration profile calculated from Eq. \ref{eq: c(zeta,t) step diffusion}, which gives $\tau=\frac{\pi D}{h} \simeq 2.3 \, 10^{3}$ s.

\section*{Discussion and Conclusions}
In this work, we have used a commercial microscope and a custom sample cell derived from a commercial optical lens holder to perform accurate scattering measurements on a critical mixture above the critical temperature $T_{c}$, with the system in one phase, and below, with the system in two phases. In the first case, the mixture was studied at different temperatures by approaching $T_{c}$ and the critical scaling of the scattering intensity and of the diffusion coefficient were obtained with good accuracy. In the second case, a sudden temperature change triggered the onset of NCF, whose interplay with the underlying critical ECF was studied during the diffusive relaxation process leading to the new equilibrium condition that follows the temperature change. During the transient, the diffusion coefficient was not constant, exhibiting an exponential relaxation compatible with the kinetics of the concentration profile. Similarly, also the equilibrium intensity $I_{eq}$ as well as the crossover wave-vector $q_{co}$ reached their new values in an exponential manner with the same time constant. Future work will be aimed to expanding the wave-vector range to access also wave-vectors smaller than $q_{ro}$. Our setup, characterised by a limited depth of field, could be also used to perform a thorough characterisation of the mixture critical properties in the two-phase region, by focusing separately on each one of the two phases and also on the interface between them. This study was not possible here because the long times of equilibration are not compatible with the short life of the sample that degrades after a few days.

%
%

%
%
%

\section*{Acknowledgments}
We acknowledge funding by the Italian Ministry of Education and Research, ‘‘Futuro in Ricerca’’ Project ANISOFT (RBFR125H0M).

\bibliographystyle{ieeetr}
\bibliography{biblio}

%
%

\end{document}